\documentclass{IEEEtran}
\usepackage{cite}
\usepackage{amsmath,amssymb,amsfonts}
\usepackage{algorithmic}
\usepackage{graphicx}
\usepackage{textcomp}
\usepackage{color,soul}
\usepackage{enumitem}
\usepackage[numbers]{natbib}
\usepackage{booktabs}
\usepackage{float}
\usepackage{multicol}
\usepackage{subcaption}
\usepackage{comment}
\usepackage[pdfborder={0 0 0}]{hyperref}

\newcommand{\Real}{\mathbb{R}}

\newcommand{\zv}{{z}}
\newcommand{\wv}{{w}}
\newcommand{\Am}{{A}} 
\newcommand{\Pm}{{P}}
\newcommand{\Sm}{{S}}
\newcommand{\im}{{I}}
\newcommand{\Mm}{{M}}
\newcommand{\Wm}{{W}}
\newcommand{\Zm}{{0}}
\newcommand{\Lambdam}{{\Lambda}} 
\newcommand{\Psim}{{\Psi}}
 
\newcommand{\Qm}{{Q}}

\newtheorem{proposition}{Proposition}
\newtheorem{remark}{Remark}

\newenvironment{Proof}[1][Proof]{\textbf{#1.} }{\ \rule{0.5em}{0.5em}}

\definecolor{MyBlue}{rgb}{0, 0, 0}
\definecolor{MyBlueR}{rgb}{0, 0, 0}
\definecolor{myviolet}{rgb}{0.55, 0.0, 0.55}
\definecolor{MinorChange}{rgb}{0, 0, 1}
\definecolor{Reviewer}{rgb}{0.55, 0.0, 0.55}
\definecolor{ComplicatePhrase}{rgb}{0.0, 0.72, 0.92}
\definecolor{PossibleToDelete}{rgb}{0.45, 0.31, 0.59}

\newcommand{\jesus}[1]{\textcolor{MyBlue}{#1}}
\newcommand{\jesusR}[1]{\textcolor{MyBlueR}{#1}}

\newcommand{\discussion}[1]{\textcolor{MyBlue}{#1}}
\newcommand{\lastmodif}[1]{\textcolor{MyBlue}{#1}}

\newcommand*{\mm}{%
  \leavevmode
  \hphantom{0}%
  \llap{%
    \settowidth{\dimen0 }{$0$}%
    \resizebox{1.1\dimen0 }{\height}{$-$}%
  }%
}

\def\BibTeX{{\rm B\kern-.05em{\sc i\kern-.025em b}\kern-.08em
    T\kern-.1667em\lower.7ex\hbox{E}\kern-.125emX}}
\begin{document}
\title{Turbidity Control in Sedimentation Columns by Direction Dependent Models}
\author{Jesus-Pablo Toledo-Zucco, Daniel Sbarbaro, João Manoel Gomes da Silva Jr.
\thanks{J-P. Toledo-Zucco is with the Département de Traitement de l'information et systèmes (DTIS) at the ONERA, the French Aerospace Lab, Toulouse 31000, France (e-mail: jtoledoz@onera.fr).}
\thanks{D. Sbarbaro is with the Departamento de Ingeniería Eléctrica at the Universidad de Concepción, Concepción 4070386, Chile (e-mail: dsbarbar@udec.cl).}
\thanks{J. M. Gomes da Silva Jr. is with the Departamento de Sist. Elétricos de Automação e Energia (DELAE) at the Universidade Federal do Rio Grande do Sul, Porto Alegre-RS 90035-051, Brasil (e-mail: jmgomes@ufrgs.br).}
\thanks{This work was partially supported by Project AMSUD220013  and Fondap SERC 15110019.}
}
\maketitle

\begin{abstract}
.
{Sedimentation is a crucial phenomenon in recovering water from slurries by separating solid-liquid. Thickeners and sedimentation columns are equipments widely used in the process industry to reclaim water from process slurries. This contribution addresses the problem of controlling the turbidity of the recovered water in a sedimentation column by manipulating the underflow. The phenomenological model describing the turbidity is too complex to be used in a control strategy, and it is difficult to identify its parameters using plant measurements. This work proposes an empirical piece-wise time-delay model for modeling the turbidity at the top of the column to circumvent these problems. A systematic design procedure is developed to tune a Proportional Integral controller guaranteeing closed-loop stability for systems modeled as a piece-wise time delay model. Experiments in a pilot plant validate the theoretical results and illustrate the control performance under various operational scenarios.}

\end{abstract}


\section{Introduction}
\label{sec:introduction}

 Reducing fresh water consumption is a major goal in the mining industry to ensure sustainability. Solid-liquid separation plays a key role in all the water recovering processes \citep{Concha2014BookSolid}. The water can be recovered from the mining slurry using a sedimentation process. In this context, it is fundamental to control the turbidity of the water recovered for its reuse in the process. As pointed out in \citep{BETANCOURT201334} and \citep{Diehl2008}, it is challenging to stabilize this control loop, and therefore a simple mass balance controller has been proposed instead. Still, these strategies require an estimation of the solid inventory. 
 
The sedimentation process can be modeled using conservation laws leading to a model described by nonlinear partial differential equations \citep{Concha2014BookSolid}\citep{BURGER2007274}\citep{LANGLOIS2019131}. This complex model can not be used in a control strategy, and its parameters are not easily identified using operational data. Tan et al. \citep{TAN20151} proposed a Kalman filter to identify key sedimentation parameters of the solid material in the slurry. They used this information to tune a model predictive control based on a linearized model. This work, however, proposes an empirical representation to model the turbidity at some specific height of the sedimentation equipment. In particular, since the solid-liquid separation process exhibits some direction-dependent dynamics is intuitive to model the complete process with two finite-dimensional models that are piece-wise depending on the direction of the variables. \jesus{The main complexity of the linear piece-wise model considered in this work is that the input delay is also switching on time.}  

Direction-dependent systems can be found in many industrial processes such as heating or cooling and chemical ones, among others \citep{Goodfrey1972}\citep{Goofrey1974}\citep{Tan2009}. These type of systems can be regarded as a special type of linear switching systems; where the switching function is defined by the process itself. \jesus{Gain-scheduled PID control with two sets of parameters has been proposed to control this type of process, along with auto-tuning methods \citep{Baskys2006ConferenceAsymmetric}\citep{Baskys2010ConferenceControl}\citep{Tan1998JournalAutomatic}\citep{Wang1998JournalAutomatic}. Although these design approaches have proven to be effective, the closed-loop stability of the switched system has not been systematically addressed.}

\jesus{On the other hand, most standard industrial PID controller blocks do not offer the capability to adjust the controller parameters based on external measurements. Tuning a PID controller in this scenario is typically performed by considering worst-case conditions. Therefore, the goal of this work is to propose a systematic approach to ensure closed-loop stability and achieve a consistent level of performance with a single tuning parameter. This approach explicitly takes into account directional-dependent dynamics (i.e., a switching behavior) and ensures a specific level of performance, expressed in terms of a guaranteed quadratic cost.}


The stability of linear switching system not only depends on the structure and the parameters of the different modes, but also on the switching function. The necessary and sufficient conditions for asymptotic stability of systems with arbitrary switching  are surveyed in \citep{Lin2009}. The analysis is centered in finding a Common Quadratic Lyapunov Function (CQLF) for establishing the closed loop stability \citep{Leith2003}.  The use of a CQLF in the context of control design for switched linear systems have been addressed by several authors. For instance, in \citep{Chaib2006} a method to design dynamic controllers for continuous time switched linear system is provided. In \citep{Chen2014} the problem of designing a switching signal for exponential stabilization of a discrete-time switched system with time-varying delay is addressed. By applying a Common Lyapunov-Krasovskii  Functional (CLKF), sufficient conditions to guarantee the global exponential stability of the closed loop system are obtained.  It is important to notice that these results can not directly be applied to the proposed sedimentation model, since its time delay is direction dependent. In order to address this problem, in our previous work \citep{Toledo2018ConferenceTuning}, a simple transformation was proposed in order to define a CLKF. The Lyapunov-Krasovskii candidate function is similar to the one proposed in \citep{Chen2003} and \citep{Chen2005}. Based on this analysis, 
 a tuning methodology for a PI controller guaranteeing closed-loop stability was proposed. In this contribution, the theoretical results are summarized for sake of completeness and applied to a sedimentation pilot plant to shade light on their applicability to real systems.

The paper is organized as follow. Sections II describes the characteristics of the sedimentation column. In section III, the identification of an empirical model is addressed. Section IV  derives the necessary conditions for closed-loop stability  and section V proposes a tuning strategy for a PI controller. Section VI illustrates the application of the proposed methodology, and presents the experimental results obtained in a sedimentation column. Finally, section VII summarizes the main conclusions.

\section{Description of the system} \label{Sect:Description}
\begin{figure}
\begin{subfigure}{.25\textwidth}
\hspace{0.8cm}
  \includegraphics[height=1.7\linewidth]{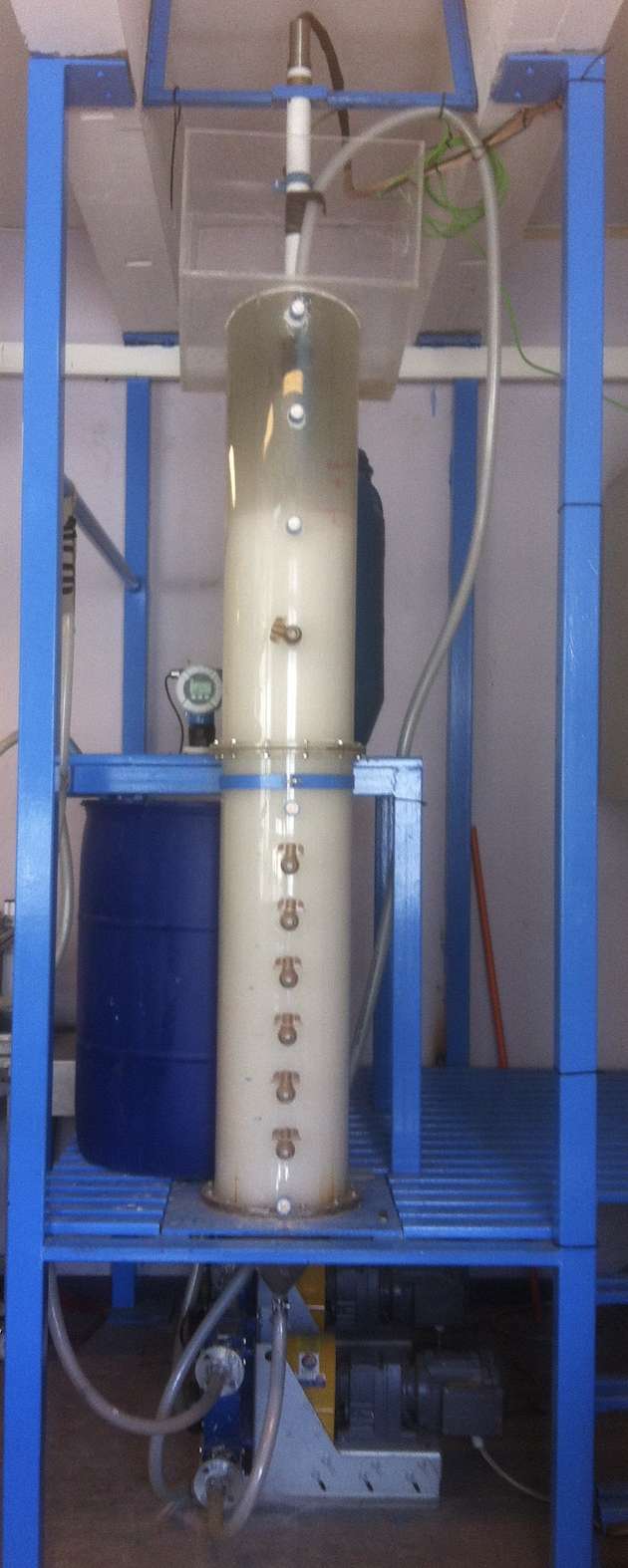} 
  \caption{ }
\end{subfigure}%
\begin{subfigure}{.25\textwidth}
  \includegraphics[height=1.8\linewidth]{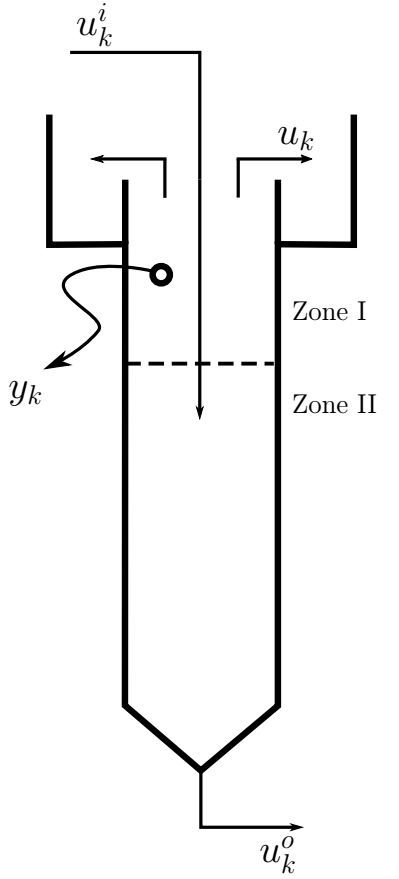} 
  \caption{ }
\end{subfigure}\\%
\begin{subfigure}{.5\textwidth}
  \centering
  \includegraphics[width=0.6\linewidth]{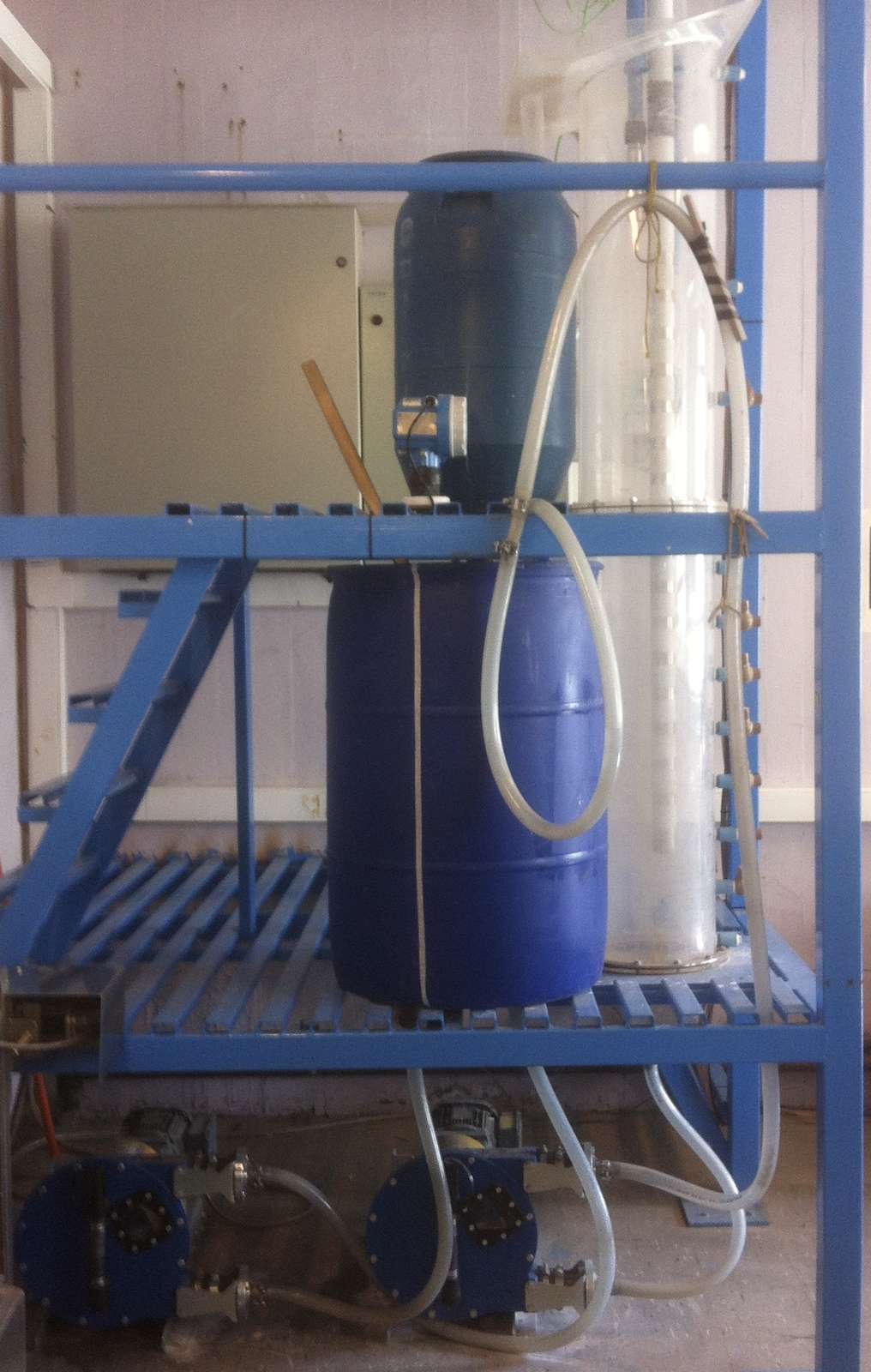} 
  \caption{ }
\end{subfigure}
\caption{\jesus{(a) Front view of the set up, (b) Schematic diagram of the process, and (c) Lateral view of the set up.}}
\label{Fig:ExperimentalSetup}
\end{figure}

The experimental setup shown in Figure \ref{Fig:ExperimentalSetup} was built for testing modelling and control strategies under wide range of operational conditions \citep{Lira2013ThesisMonitoreo} \citep{Toledo2015ThesisModelacion}.  The sedimentation column is  $2.5\, m$ height and has a $0.5 \, m^2$ cross-section surface. In figure \ref{Fig:ExperimentalSetup} a),  the frontal side of the sedimentation column in a normal operation shows two different zones:
\begin{enumerate}[label=\roman*.]
\item Zone 1: Clarification zone at the top part of the sedimentation column.
\item Zone 2: Thickening zone at the bottom part.
\end{enumerate}
Zone 1 has a small amount of solid particles compared to Zone 2. The objective is to keep a certain overflow at the top of the column with a fixed water turbidity. The recovered water is stored in a container for re-utilization. As seen in Figure \ref{Fig:ExperimentalSetup} c), a second tank with a mixer is used to store the slurry and to pump it to the sedimentation column.
Two pumps are used to control the feed and discharge flows. Both pumps are commanded by variable frequency drives. They are located under the column, as seen in \ref{Fig:ExperimentalSetup} c). Note that, having an input flow greater than the output one, an overflow can be generated at the top of the sedimentation column. Finally, at Zone 1 a turbidity sensor measures the turbidity of the overflow (see  Figure \ref{Fig:ExperimentalSetup} a) and the scheme of Figure \ref{Fig:ExperimentalSetup} b)).

The main variables of the experimental set-up are the frequencies of the feed pump drives for the feed flow $u^i_k$, and discharge flow $u^o_k$, and the turbidity at the top of the column $y_k$ (see Figure \ref{Fig:ExperimentalSetup} b)). Variable $k$ represents the sample at time $t = kT$, with $T$ denoting the sampling time of the controller. All the instrumentation is connected to an Allen Bradley Control Logic PLC. We define the control input as $u_k = u^i_k - u^o_k$. In the following section, we propose an empirical piece-wise time-delay model for modelling the turbidity $y_k$ as a function of the control input $u_k$.

%

\section{Empirical Model} \label{Sect:EmpModel}
\begin{figure}[thpb]
  \centering
    \includegraphics[width=0.4\textwidth]{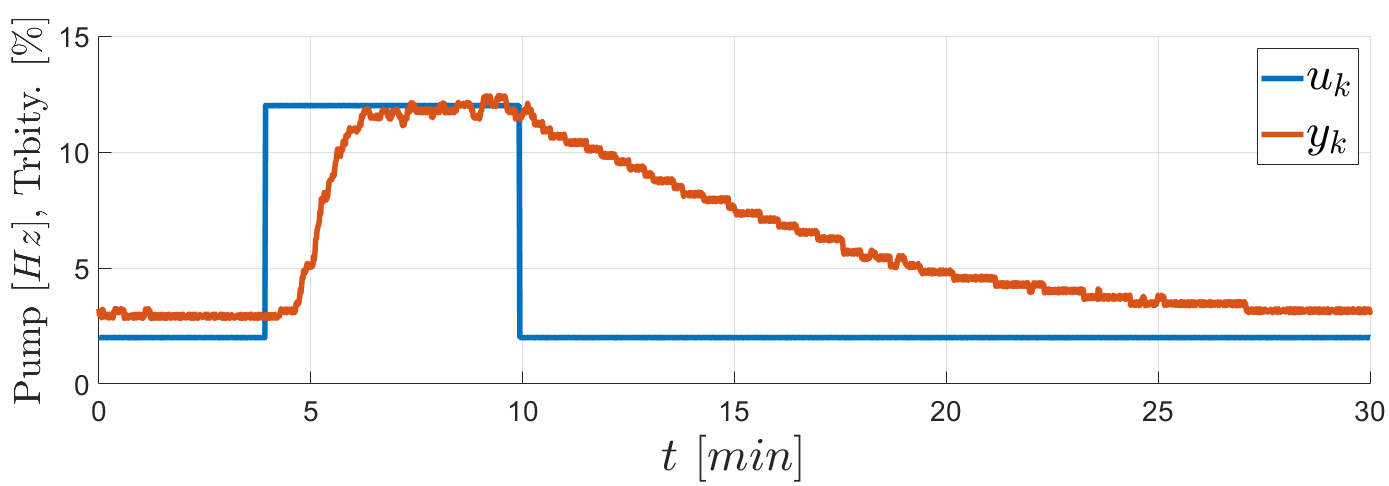}
   \caption{Output response for a step input}
   \label{Fig:OpenLoopResponse}
\end{figure}

\begin{figure}[thpb]
  \centering
    \includegraphics[width=0.4\textwidth]{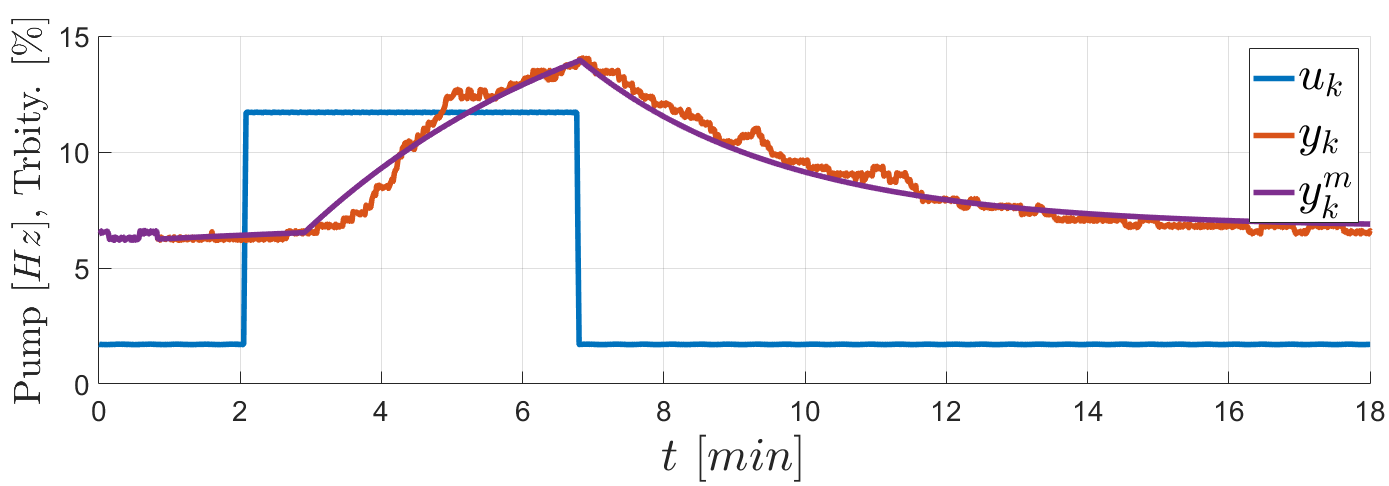}
   \caption{Comparison between the model $y_k^m$ and the real values of $y_k$, and \jesus{the input signal $u_k$}.}
   \label{Fig:ModelVsData}
\end{figure}

 The model is identified using the measured data of the turbidity when perturbing the column with step changes. As seen in Figure \ref{Fig:OpenLoopResponse}, for a step inputs the output behaves as a first order system with the following characteristics:
\begin{enumerate}[label=\roman*.]
\item The dynamic of $y_k$ behaves with different time constants, when $u_k$ is increasing ($\Delta u_k := u_k - u_{k-1} >0$) than when it is decreasing ($\Delta u_k <0$). 
\item The delay is different when $u_k$ is increasing ($\Delta u_k >0$) than when it is decreasing ($\Delta u_k <0$).
\item The output $y_k$ has a saturation limit.
\end{enumerate}
For simplicity, we consider that the measured turbidity does not reach the saturation limit. Hence it can be modeled as a piece-wise or direction dependent system \citep{Tan2009}. The dynamic model writes
\begin{equation} \label{Eq:Model}
\begin{split}
x_{k+1} &= a_{\sigma} x_k + b _{\sigma} u_{k-d_{\sigma}}+ c_{\sigma},\\
y_k &= x_k
\end{split}
\end{equation}
\jesus{The variable $\sigma$ denote the direction mode at each instant $k$ as follows:
\begin{equation}
\begin{split}
\jesus{{\sigma}(k)} & =\,\,\,\left\{
\begin{array}{ll}
      1, & \hspace{1cm} {\Delta u}_k>0, \\
      2, & \hspace{1cm} {\Delta u}_k<0, \\
      \jesus{{\sigma}({k-1})}, & \hspace{1cm} {\Delta u}_k=0, \\
\end{array} 
\right.
\end{split}
\end{equation}
For notation simplicity, we drop the $\sigma$ time-dependence in \eqref{Eq:Model}.} For $i= \lbrace 1,2 \rbrace$, $a_i\in \mathbb{R}$ is proportional to the time constant, $b_i\in \mathbb{R}$ is the input gain, $c_i \in \mathbb{R}$ is an offset value, and $d_i \in \mathbb{N}$ is the time delay. Notice that the model's delay is also direction dependant. \jesus{Several open-loop experiments were carried out to gather information of the process dynamics}. The parameters are obtained by solving a least square optimization problem. Figure \ref{Fig:ModelVsData} shows the process and model responses using a validation data set \jesus{with a sampling period $T_s = 1\,s$}. The identified parameters are summarized in Table \ref{Tab:ParameterSim}.

\begin{table}[thpb]
\centering
\caption{Identified experimental parameters}
\label{Tab:ParameterSim}
\begin{tabular}{lllll}
            & 		$a_\sigma$ & $b_\sigma$ & $c_\sigma$ & $d_\sigma$ \\
$\sigma=1$	& 0.9962      	& 0.0046    & 0.0189 &   50 \\
$\sigma=2$	& 0.9942      	& 0.0084    & 0.0245 & 1 \\
\end{tabular}
\end{table}
{In the following, we propose systematic design procedure to tune a PI controller so that the closed-loop stability is guaranteed.}

\section{Controller design} \label{Sect:ContrDesign}
\begin{figure}[thpb]
\vspace{-0.5cm}
  \centering
    \includegraphics[width=0.4\textwidth]{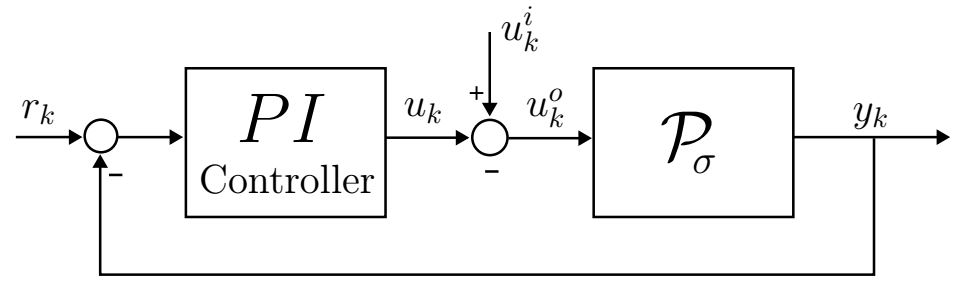}
   \caption{Block diagram of the closed-loop system.}
   \label{Fig:BlockDiagram}
\end{figure}

The design of a Proportional-Integral (PI) controller for the system \eqref{Eq:Model} such that the closed-loop system is asymptotically stable was proposed in \citep{Toledo2018ConferenceTuning}. This approach finds a region for the controller parameters in which the closed-loop stability is guaranteed and then selects a single pair of values  to minimize the settling time of $x_k$. The controller gains are designed by means of an iterative solution of a set of Linear Matrix Inequalities (LMIs). \jesus{In Fig. \ref{Fig:BlockDiagram}, we show the block diagram of the closed-loop system, consisting of a classic output feedback $y_k$, compared to the reference signal $r_k$, passing through a PI controller and obtaining the control signal $u_k$. Finally, the frequency of the output pump drive $u_k^o$ is obtained subtracting the frequency of the input pump drive $u_k^i$ from the control signal, {\it i.e.,} $u_k^o = u_k^i - u_k$.}

For the system \eqref{Eq:Model}, we define the the reference signal $r_k$, the error $e_k = r_k - y_k$, and the PI controller:
\begin{equation} \label{Eq:ControlLaw}
u_k = k_pe_k + k_i i_k, \quad i_k := \sum_{j=0}^{k-1} e_j ,
\end{equation}
 in which $k_p  \in \mathbb{R}$ and $k_i  \in \mathbb{R}$ are the proportional and integral gains, respectively. 
 
In the following, two propositions summarize the main theoretical results of this paper. Using these results, one can verify if a pair $(k_p,k_i) \in \Real ^2$ is able to stabilize the system \eqref{Eq:Model} and search for the pair $(k_p,k_i)$ that minimizes a \jesus{quadratic} cost function related to the settling time of the turbidity. \jesus{In the following, since the bias terms $c_1$ and $c_2$ are compensated by the integral action of the controller in both operational modes, we omit them for the stability analysis in order to simplify the explanation of the following propositions.} 
Without loss of generality we also consider a null reference signal $r_k = 0$ for all $k$. 

\begin{proposition}\label{Prop:Main1} {Define the following matrices for the open-loop system \eqref{Eq:Model} and the control law \eqref{Eq:ControlLaw}: 
\begin{equation}
{A}_{ \sigma  }=\begin{pmatrix}
 {a}_{ \sigma  } & 0 \\
1 & 1
 \end{pmatrix},\quad
{B}_{ \sigma }=\begin{pmatrix}
 {b}_{ \sigma  }   \\
0
 \end{pmatrix}, \quad
 {C} =\begin{pmatrix}
 {1} & 0 \\
0 & 1
 \end{pmatrix},
\end{equation}
\begin{equation}
   \sigma \in \lbrace 1,2 \rbrace , \quad  K_c = \begin{pmatrix} k_p & k_i \end{pmatrix}.
\end{equation}} The closed-loop system \eqref{Eq:Model}-\eqref{Eq:ControlLaw} is asymptotically stable if there exist $4\times 4$ matrices $ P_1>0_4 $, $ S_1>0_4 $, $ S_2>0_4$, $ W_1 $, $W_2 $, $W_3 $, $M_1 $, $M_2 $, $P_2 $, and $P_3 $ such that the following LMIs are satisfied:
\begin{equation}\label{Ineq:LMIS}
\begin{split}
    \Lambda & < 0_{12} , \\
    \widetilde{A}_2^T P_1 \widetilde{A}_2 - P_1 &< 0_4, \quad 
    {\Lambda_2} {< 0_4},\\
    \begin{bmatrix}
    W & M\\
    M^T & S_1
    \end{bmatrix} &\geq 0_{12},
\end{split}
\end{equation}
with 
\begin{equation*}
\Lambdam =\left[ \begin{matrix} \Psim  & { \Pm }^{ T }\left[ \begin{matrix} \Zm_4  \\ \widetilde{ \Am }_{ p } \end{matrix} \right] \mm \Mm \\ \left[ \begin{matrix} \Zm_4  & \widetilde{ \Am }_{ p }^{ T } \end{matrix} \right] \Pm \mm { \Mm }^{ T } & { \mm \Sm }_{ 2 } \end{matrix} \right], 
\end{equation*}
\begin{align*}
\Psim=&  h\Wm+\begin{bmatrix} { \Sm }_{ 2 } & \Zm_4  \\ \Zm_4  & { \Pm }_{ 1 }+h{ \Sm }_{ 1 } \end{bmatrix} +\left[ \begin{matrix} \Mm & \Zm_{8\times 4}  \end{matrix} \right] +\left[ \begin{matrix} \Mm^{ T } \\ \Zm_{4\times 8}  \end{matrix} \right] \\ 
+& { \Pm }^{ T }\begin{bmatrix} \Zm_4  & \im_4 \\ \widetilde{ \Am }_1 \mm \widetilde{ \Am }_p \mm \im _4& \mm\im_4 \end{bmatrix}+\begin{bmatrix} \Zm_4  & \widetilde{ \Am }_1^{ T} \mm \widetilde{ \Am }_p^{ T } \mm \im_4 \\ \im_4 & \mm \im _4\end{bmatrix}\Pm,
\end{align*}
\begin{equation*}
    W = \begin{bmatrix}
    W_1 & W_2 \\ W_2^T & W_3
    \end{bmatrix}, \quad M = \begin{bmatrix}
    M_1  \\ M_2 
    \end{bmatrix}, \quad P = \begin{bmatrix}
    P_1 & 0_4 \\ P_2 & P_3
    \end{bmatrix},
\end{equation*}
\begin{equation*}
    \widetilde{A}_1 = \begin{bmatrix}
    A_1 & \mm B_1 K_c C \\
    I_2 & 0_2
    \end{bmatrix}, \quad \widetilde{A}_2 = \begin{bmatrix}
    A_2 & \mm B_2 K_c C \\
    I_2 & 0_2
    \end{bmatrix} \end{equation*}
    \begin{equation*}\widetilde{A}_p = \begin{bmatrix}
    0_2  & \mm B_1 K_c C \\
    0_2 & 0_2 
    \end{bmatrix},\quad h = d_1-d_2,
\end{equation*}
{\begin{align*}
\Lambda_2 &= h (\widetilde{A}_2^{T})^h \tilde{S}_1 (\widetilde{A}_2)^{h} - \sum _{i = 0} ^{h-1} \widetilde{A}_2 ^{i ^ T} \tilde{S}_1 \widetilde{A}_2 ^{i}  \\ &  \quad \quad \quad \quad \quad \quad \quad \quad \quad \quad \quad \quad +(\widetilde\Am^T_2)^h { \Sm }_{ 2 }(\widetilde\Am_2)^h 
 -{ \Sm }_{ 2 },\\ 
\tilde{S}_1 &= (\widetilde{A}_2 - I)^T S_1 (\widetilde{A}_2 - I).
\end{align*}}
\end{proposition}
\begin{Proof}
See Appendix \ref{Proof1}.
\end{Proof}

{
\begin{remark}
    The $\im_i$ notation refers to the identity matrix of size $i$, the $\Zm_i$ to a null square matrix of size $i$, and $\Zm_{i\times j}$ to a null matrix of $i$ rows and $j$ columns.
\end{remark}}
\begin{remark}
With the previous proposition one can find a region $\Omega_1$, such that $(k_p,k_i) \in \Omega_1 \subset \Real^2$ assures the closed-loop stability. This region can be found by iteratively solving the LMI \eqref{Ineq:LMIS} for a mesh grid of pairs $(k_p,k_i)$. Since $d_2 = 1$, the unknown matrices of the LMI are of size $4\times 4$. If $d_2$ is bigger, then the unknown matrices are of bigger size as well.
\end{remark}

\jesus{\begin{remark}
The conditions in \eqref{Ineq:LMIS} are obtained to guarantee stability of both models with a common Lyapunov-Krasovskii functional. The inequality $\Lambda < 0$ guarantees the stability when $\sigma =1$. We use the Moon's inequality \citep{Chen2003}, which lead to the forth LMI in \eqref{Ineq:LMIS} (see \eqref{eq:LMI4} in the proposition proof in Appendix \ref{Proof1}). The inequalities $\widetilde{A}_2 ^\top P_1 \widetilde{A}_2 - P_1 <0$ and $\Lambda_2<0$ guarantees closed-loop stability when $\sigma = 2$ (see the derivation of \eqref{eq:LMI6} in Appendix \ref{Proof1}).
\end{remark}
}

{For design purposes, the LMIs in \eqref{Ineq:LMIS} are slightly modified in such a way that the closed-loop system remains asymptotically stable and minimizes the following cost function:
\begin{equation}\label{Eq:CostFunction}
    J = \sum_{k=0}^{\infty} z_k^T Q z_k, \quad with \quad  z_k = \begin{bmatrix}
    x_k \\
    - i_k \\
    x_{k-1} \\
    - i_{k-1} \\
    \end{bmatrix},
\end{equation}
where $Q \in \Real ^{4 \times 4}$ is a design matrix that can weight the current state ($x_k$), the current integral action ($-i_k$) and the delayed ones by the smallest delay $d_2$, which in this case is unitary.
}

\begin{proposition}\label{Prop:Main2}
Consider the same conditions and matrices than in Proposition \ref{Prop:Main1}. Replace the LMIs in \eqref{Ineq:LMIS} by the following:
\begin{equation}\label{Ineq:LMIS2}
\begin{split}
    \Lambda & < -\overline{Q} , \\
    \widetilde{A}_2^T P_1 \widetilde{A}_2 - P_1 &< -Q, \quad 
    {\Lambda_2}  < {0_4},\\
    \begin{bmatrix}
    W & M\\
    M^T & S_1
    \end{bmatrix} &\geq 0_{12}, \quad \overline{Q} =  \begin{bmatrix} Q & 0_4 & 0_4\\0_4 & 0_4 & 0_4\\ 0_4 &0_4 & 0_4\end{bmatrix},
\end{split}
\end{equation}
with $0_4\leq Q= Q^T \in \Real^{4 \times 4}$ and $\overline{Q} \in \Real^{12 \times 12}$. If we select $(k_p,k_i)$ such that \eqref{Ineq:LMIS2} is satisfied and the trace of the matrix $P_1+hS_2$ is minimum, \discussion{then a minimized guaranteed cost is obtained, that is $J < z_0^\top (P_1 + h S_2) z_0$, considering that $z_k=z_0$, $\forall k \in [-h,0]$. } 
\end{proposition}

\begin{Proof}
See Appendix \ref{Proof2}.
\end{Proof}
\begin{remark}\label{Remark:Q}
The minimization of the trace of the matrix $P_1 +hS_2$ can be solved iteratively by evaluating the trace of $P_1 +hS_2$ at every pair $(k_p,k_i) \in \Omega_1$. A simple choice of $Q$ can be a diagonal matrix with the two first diagonal terms
weighing the plant state ({\it i.e., the output}) and the integral term, respectively.
\end{remark}

\jesus{\begin{remark}\label{Rem:Gain-Scheduling}
\jesusR{A gain-scheduling control approach can be used to optimize performance in both directions ($\sigma = \lbrace 1,2 \rbrace$), potentially yielding superior results compared to those proposed in this article. However, this method requires tuning four
control parameters (a pair of $(k_p,k_i)$ per model). For the sake of simplicity, we have opted for a single PI controller capable of stabilizing the models of both modes.}
\end{remark}}

{\jesus{It should be noted that the LMIs in \eqref{Ineq:LMIS} can also be used to certify the closed-loop stability if other methods are used to tune the PI parameters.} In the following section, we propose a brief design procedure to tune the gains of the PI controller based on the previous results.}


\section{Design Procedure} \label{Sect:Procedure}

{The design procedure can be summarized as follows\footnote{The Matlab codes can be downloaded from \href{https://gitlab.com/ToledoZucco/tunepi}{https://gitlab.com/ToledoZucco/tunepi}} 
\begin{enumerate}
\item Select the ranges $k_p \in [\underline{k}_{p},\overline{k}_{p}]$ and $k_i\in [\underline{k}_{i},\overline{k}_{i}]$.
\item Generate a grid of pairs $(k_p,k_i)$ inside the selected ranges.
\item Solve the LMIs in \eqref{Ineq:LMIS} for every pair $(k_p,k_i)$ on the grid. If the LMIs have solution, then the pair belongs to $\Omega_1$.
\item Chose a desired weight matrix $Q$ and Solve the LMIs in \eqref{Ineq:LMIS2} for every pair $(k_p,k_i) \in \Omega_1$. If the LMIs have solution, then the pair belongs to $\Omega_2$.
\item Select the pair $(k_p,k_i) \in \Omega_2$ that minimizes the trace of the matrix $P_1+hS_2$.
\end{enumerate}}

\begin{figure}[h!]
  \centering
    \includegraphics[width=0.42\textwidth]{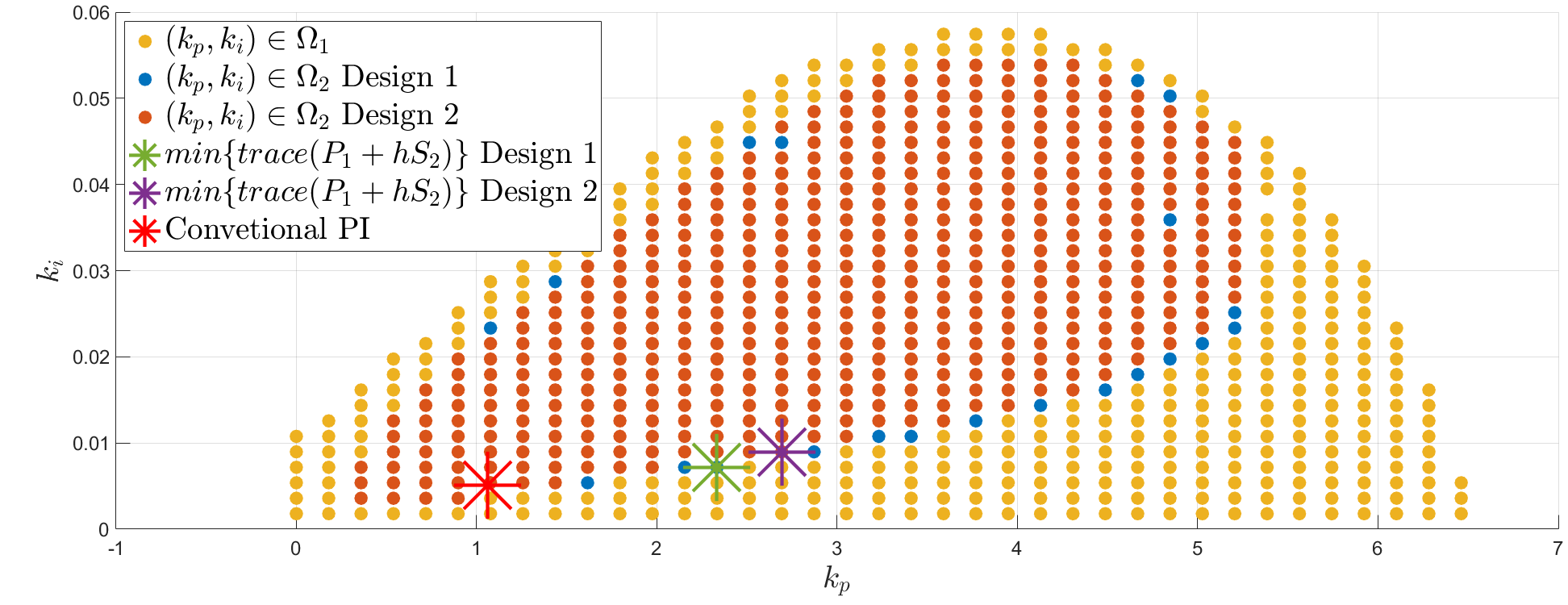}
   \caption{Set of admissible controller gains $ \Omega_1$, constrained set $\Omega_2$, and optimal pair.} 
   \label{Fig:Region}
\end{figure}

{The parameters obtained empirically in Table \ref{Tab:ParameterSim} are used to design the PI controller following the previous steps. First, we select $[\underline{k}_{p},\overline{k}_{p}] = [0,7]$ and $[\underline{k}_{i},\overline{k}_{i}] = [0,06]$. Second, we create the grid using a $40$ values per parameter. Third, we solve the LMIs in \eqref{Ineq:LMIS} for every pair $(k_p,k_i)$ on the grid. If the LMIs have solution we save the pair in the set $\Omega_1$ (see Figure \ref{Fig:Region}). Every pair in $\Omega_1$ guarantees closed-loop stability. \lastmodif{For the fourth step, we show two different designs: Design 1 with \jesusR{$Q =diag(20,20,0,0)$ and Design 2 with $Q =diag(22,22,0,0)$. By increasing the values of $Q$, one can achieve faster responses, but with bigger overshoots in the worst case (when $\sigma = 1$)}}. Then, we solve the LMIs in \eqref{Ineq:LMIS2} for every pair $(k_p,k_i) \in \Omega_1$. If the LMIs have solution, we save the pair in the set $\Omega_2$ (See Figure \ref{Fig:Region}). \lastmodif{We can see that when increasing the diagonal values of $Q$, the region $\Omega_2$ gets reduced}. Finally, we select the pair $(k_p,k_i)\in \Omega_2$ that leads to a solution to LMIs in \eqref{Ineq:LMIS2} that minimizes the trace of the matrix $P_1 + hS_2$.}
\jesus{\begin{remark}
The upper and lower limits of $k_p$ and $k_i$ can be obtained using a trial and error strategy. One way to do this is to start with $\underline{k}_p = \underline{k}_i = 0$ and \jesusR{random values for $\overline{k}_p$ and $\overline{k}_i$, and solving the LMI \eqref{Ineq:LMIS} in a short mesh. Then, increase or reduce the values of $\overline{k}_p$ and $\overline{k}_i$ accordingly. Finally, the mesh can be chosen more precise in order to extend the choices of the pairs $(k_p,k_i)$.} 
\end{remark}}

\subsection{Numerical Simulation}

\jesus{The numerical simulation that follows shows the solution characteristics obtained through the proposed approach in comparison to a conventional tuning. \lastmodif{As we show in Fig. \ref{Fig:Region}, for the Design 1 we obtain: \jesusR{$k_p = 2.3333$ and $k_i = 0.0072$}, whereas for the Design 2: \jesusR{$k_p = 2.6923$ and $k_i = 0.0090$}. \jesusR{Additionally, a conventional PI controller has been designed using the AMIGO tuning rules considering the worst case only ($\sigma = 1$).}}} \jesusR{ In this method, the controller gains are obtained as \citep[Section~2.3]{Hagglund2004JournalRevisiting}
\begin{align*}
k_p &= \tfrac{1}{K}(0.15+0.35 \tfrac{\tau}{\theta} - \tfrac{\tau^2}{(\theta+\tau)^2}), \\ 
k_i &= \tfrac{k_p}{\tau_i}, \quad \quad \tau_i = (0.35+\tfrac{6.7\tau ^2}{\tau ^2 +2 \theta \tau + 10 \theta ^2})\theta,
\end{align*}
in which $K$, $\tau$ and $\theta$ represent respectively the static gain, time constant, and time delay of a delayed first-order continuous model. We consider that these model parameters correspond to the delayed first-order model with the worst delay, {\it i.e., $\sigma = 1$ (see Table \ref{Tab:ParameterSim})}. In this case, the parameters of the continuous time model are computed as $\tau = \tfrac{1}{1-a_1}$, $K = b_1 \tau$, and $\theta = d_1$. Thus, we obtain the following controller parameters: $k_p = 1.0623$ and $k_i = 0.0051$. }

\begin{figure}[h!]
  \centering
    \includegraphics[width=0.45\textwidth]{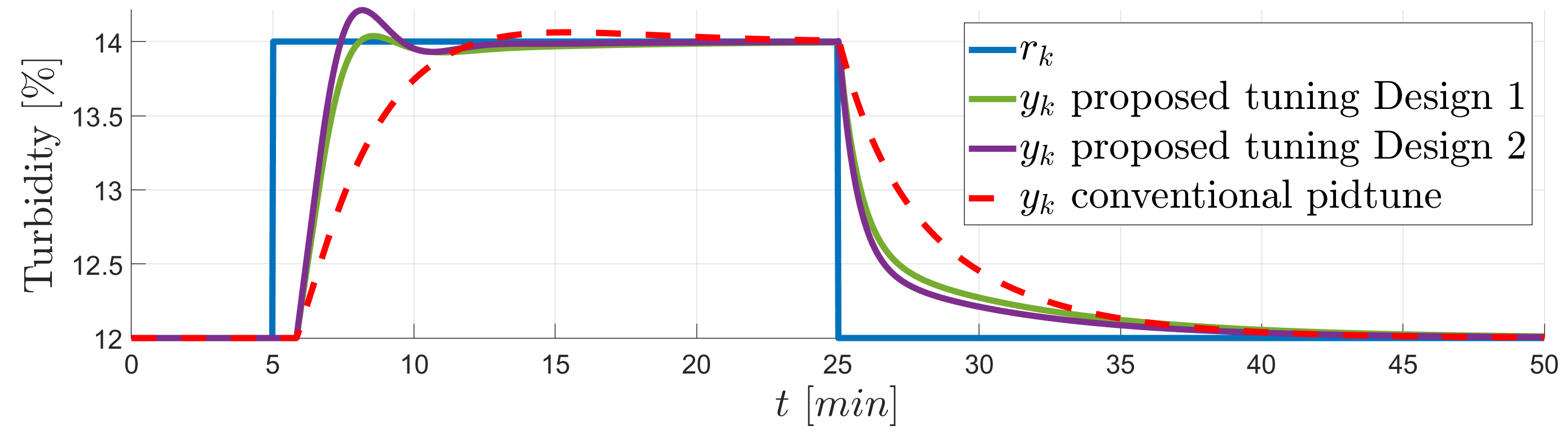}
   \caption{Reference signal $r_k$ (blue line), closed-loop responses $y_k$ using the proposed approach (\jesusR{solid green and violet lines}) and the conventional design (\jesusR{dashed red line}).} 
   \label{Fig:Direction}
\end{figure}
\begin{figure}[h!]
  \centering
    \includegraphics[width=0.45\textwidth]{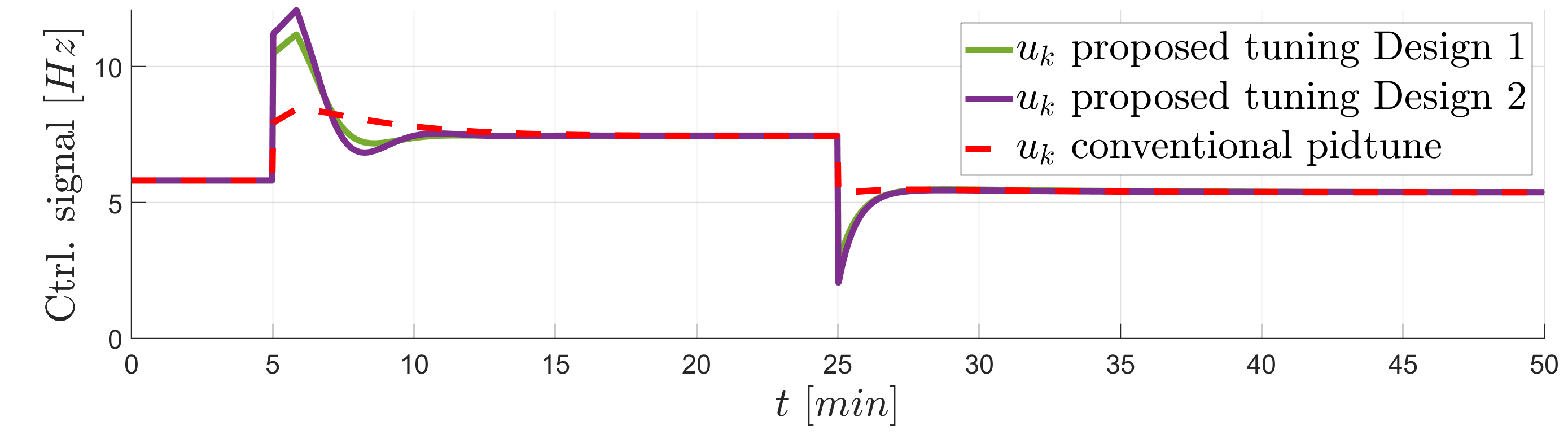}
   \caption{Control signals $u_k$ using the proposed approach (\jesusR{solid green and violet lines}) and the conventional design (\jesusR{dashed red line}).} 
   \label{Fig:InputsSwitch}
\end{figure}

 \jesus{The system \eqref{Eq:Model} with parameters summarized in Table \ref{Tab:ParameterSim}
is simulated with initial conditions $x_0 = 12\, [\%]$. In Fig. \ref{Fig:Direction} and Fig. \ref{Fig:InputsSwitch}, we show the step responses and the control signals, respectively.} \lastmodif{The conventional PI achieves a good balance between overshoot and settling time for a positive step response (see step response at $t = 5\, [min]$). However, since the second mode has not been considered in the design, the performance of negative step responses (see step response at $t = 25 \,[min]$) has a large settling time. Using the proposed approach, one can improve the behaviour of the second mode by aggravating the behaviour of the first one and always guaranteeing closed-loop stability. In this example, we can see that we can reduce the settling time for negative step responses in exchange for increasing the overshoot of the positive ones. 
As we can see in Fig. \ref{Fig:Direction} and Fig. \ref{Fig:InputsSwitch}, using the proposed method, one can reduce the overshot by modifying the weight matrix $Q$ (Design 2 has smaller overshoot than Design 1).}



\section{Experimental Results} \label{Sect:ExpResults}

\begin{figure}[!h]
\begin{subfigure}{.24\textwidth}
  \includegraphics[height=1.3\linewidth]{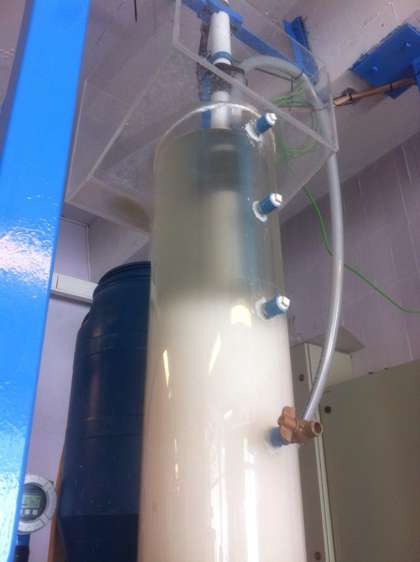} 
\end{subfigure}%
\begin{subfigure}{.24\textwidth}
  \includegraphics[height=1.3\linewidth]{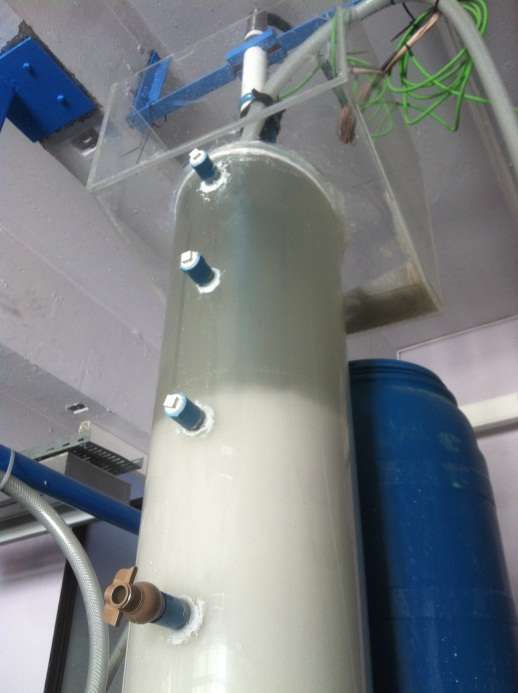} 
\end{subfigure}
\caption{At the left a turbidity $y_k \approx 1\%$ and the the right a turbidity $y_k \approx 5\%$.  }
\label{Fig:LowHighTurbidity1}
\end{figure}

\begin{figure}[!h]
\begin{subfigure}{.24\textwidth}
  \includegraphics[height=1.3\linewidth]{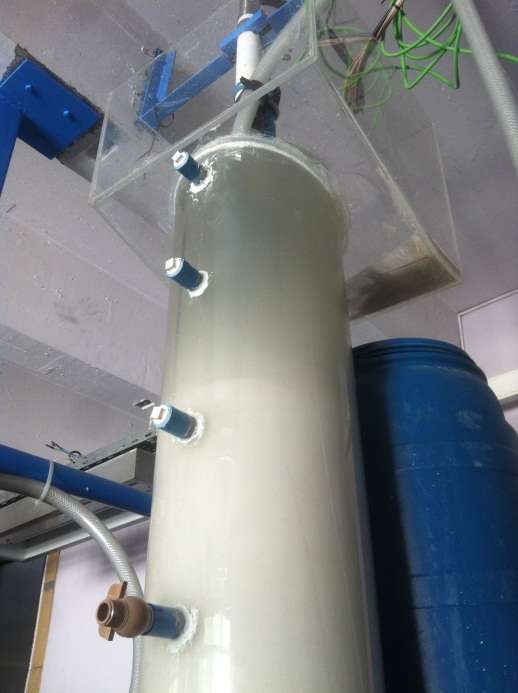} 
\end{subfigure}%
\begin{subfigure}{.24\textwidth}
  \includegraphics[height=1.3\linewidth]{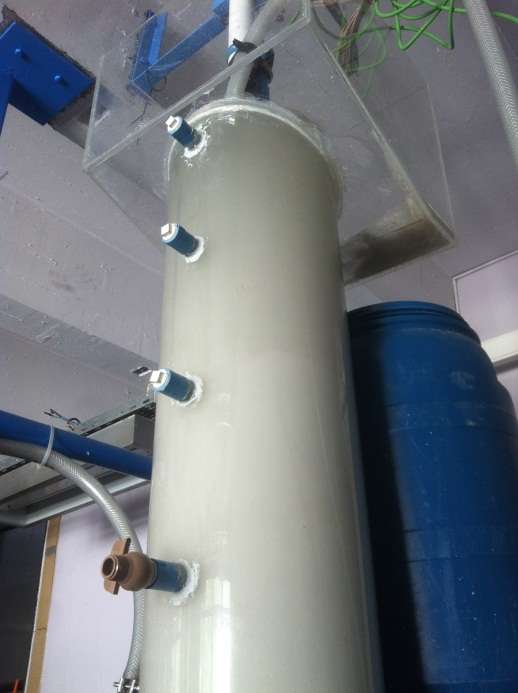} 
\end{subfigure}
\caption{At the left a turbidity $y_k \approx 10\%$ and the the right a turbidity $y_k \approx 20\%$.  }
\label{Fig:LowHighTurbidity2}
\end{figure}

The controller was tested under set-point changes and step disturbances in the inflow. Figures \ref{Fig:LowHighTurbidity1} and \ref{Fig:LowHighTurbidity2}, show the turbidity at the top of the sedimentation column for four operational points. {In the left image of Fig. \ref{Fig:LowHighTurbidity1}, we show the column with a very low turbidity at the top ($y_k \approx 1\%$). This is achieved by having a bigger outflow (at the bottom of the column) than the inflow (at the three quarters of the column). However, under these conditions there is no overflow at the top of the column for water recycling. Since the objective of this process is the water recovering, we restring ourselves to bigger values of turbidity where a small overflow is generated (see the other three images in Figs \ref{Fig:LowHighTurbidity1} and \ref{Fig:LowHighTurbidity2}) }. Using the proposed PI controller, we can set a desired turbidity and modified in real time according to water requirements. There is a trade-off between the amount of water recovered and its turbidity.


\begin{figure}[h!]
  \centering
    \includegraphics[width=0.4\textwidth]{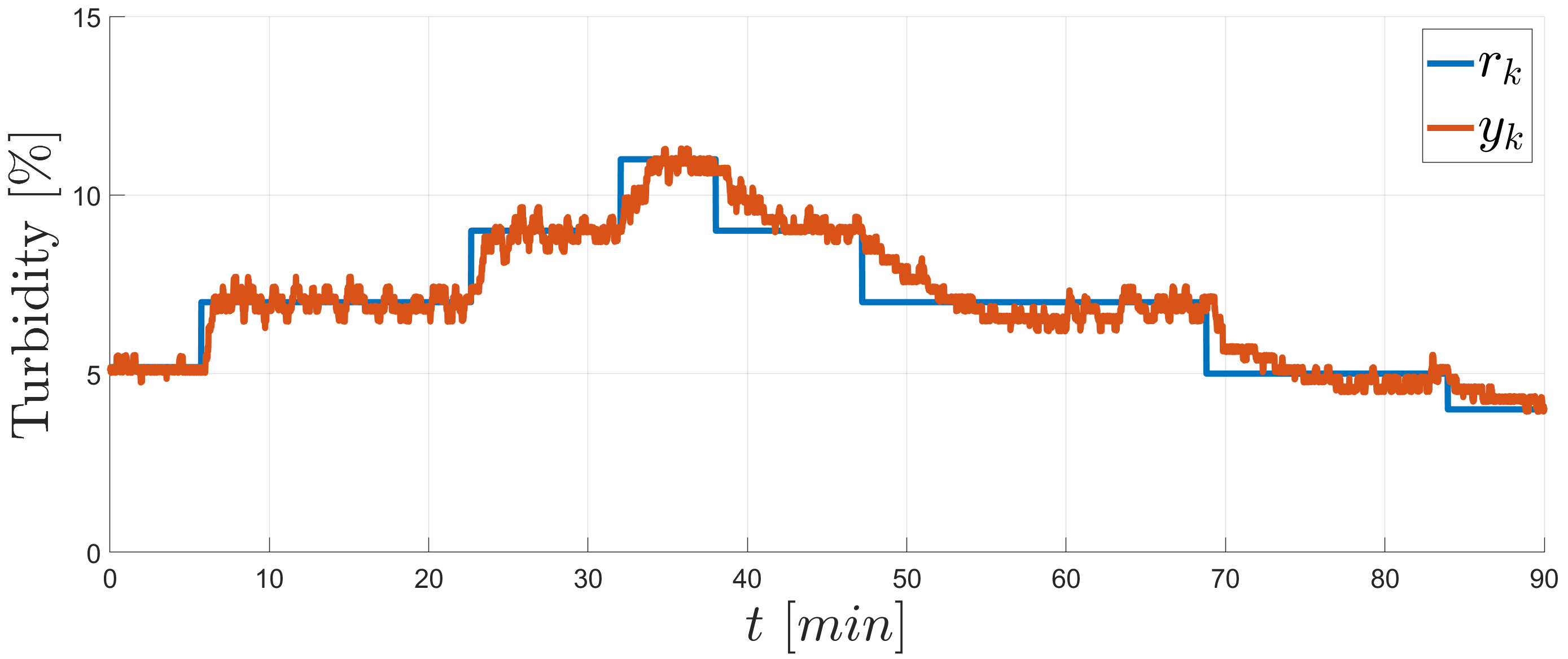}
   \caption{Closed-loop reference signal $r_k$ and turbidity $y_k$.}
   \label{Fig:ClosedLoop1}
\end{figure}

 The closed-loop piece-wise behaviour for different references values of turbidity can be seen in  Figure \ref{Fig:ClosedLoop1}. When the turbidity is increasing the closed-loop system behaves as a first order system with a constant time about $\tau \approx  1\,[min]$ and a static gain $K \approx  1$.
  When the turbidity is decreasing, the closed-loop behaviour is slower than when increasing it and it has a small overshoot. \jesus{This is mainly due to the fact that we use a single controller for stabilizing a plant that is composed of a piece-wise linear and delayed model that switches between two different models. As we point it out in Remark \ref{Rem:Gain-Scheduling}, one could consider a gain-scheduling approach in order to get an improved performance for both models.} 
\jesusR{We note that the responses depicted in Fig. \ref{Fig:ClosedLoop1} deviate from the numerical simulations presented in Fig. \ref{Fig:Direction},
primarily due to unaccounted nonlinearities and the infinite-dimensional nature of the real phenomena. At
approximately $5\%$ turbidity, which is close to the operation point for which the model has been identified, one can observe a resemblance between the experimental and numerical
responses: a rapid reaction to a positive step and a slower reaction to a negative step.}  
  
\begin{figure}[h!]
  \centering
    \includegraphics[width=0.4\textwidth]{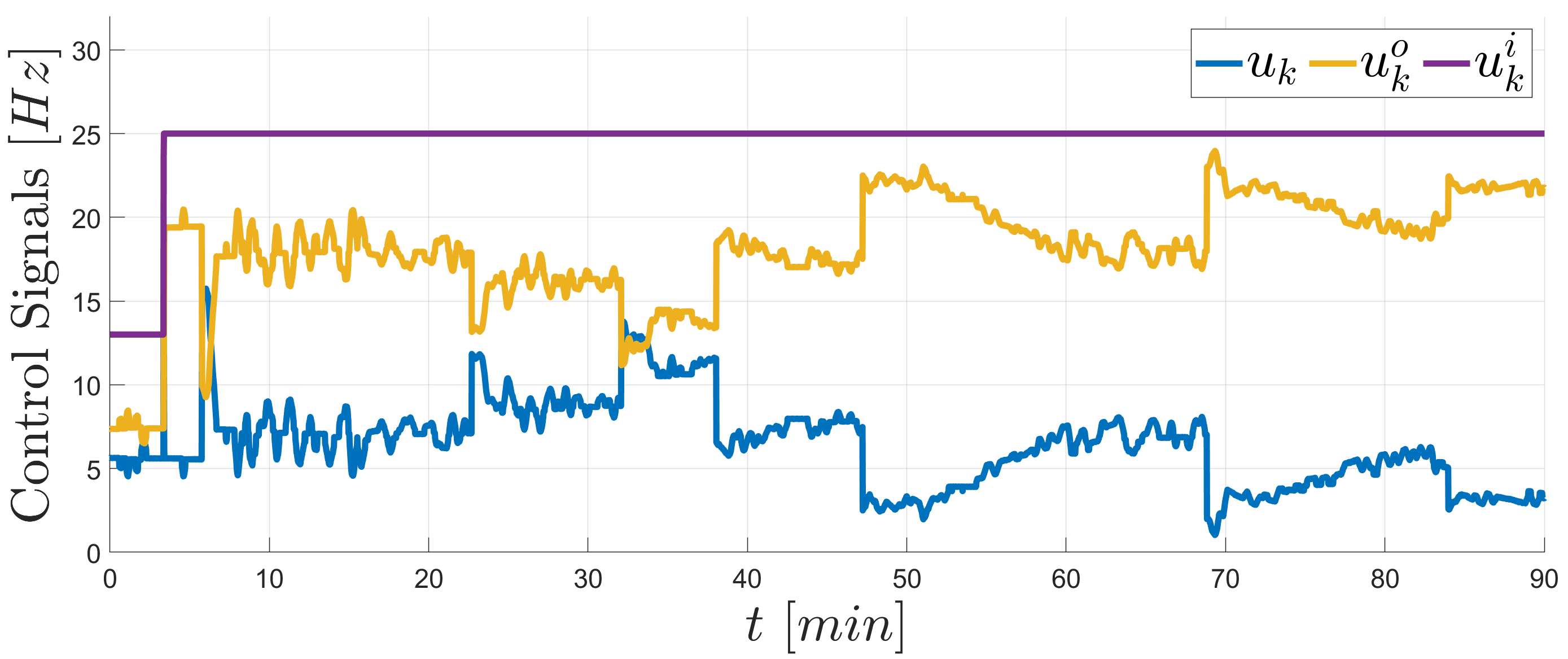}
   \caption{Closed-loop control signals $u_k$, $u_k^o$ and $u_k^i$.}
   \label{Fig:ClosedLoop1ControlSignals}
\end{figure}

  \jesus{In Fig. \ref{Fig:ClosedLoop1ControlSignals}, we show the control signal $u_k$, and the pump drive frequencies $u_k^i$ (input pump) and $u_k^o$ (output pump), where the output one is computed as shown in the block diagram of Fig. \ref{Fig:BlockDiagram}. We can see that during the test, the control signals $u_k^i$ and $u_k^o$ remains inside $[0,50]\, [Hz]$, not being saturated.}

\jesus{\begin{remark}
Notice that the noise of the output signal in Fig. \ref{Fig:ClosedLoop1} (closed-loop system) is slightly bigger than the ones in Fig. \ref{Fig:OpenLoopResponse} and Fig. \ref{Fig:ModelVsData} (open-loop system). This noise amplification may be induced by the paralell implementation of the PI controller through the proportional gain. This issue can be potentially improved by filtering the output signal before introducing it to the controller or considering a cascade implementation of the PI controller. 
\end{remark}}

\begin{figure}[!h]
  \centering
    \includegraphics[width=0.4\textwidth]{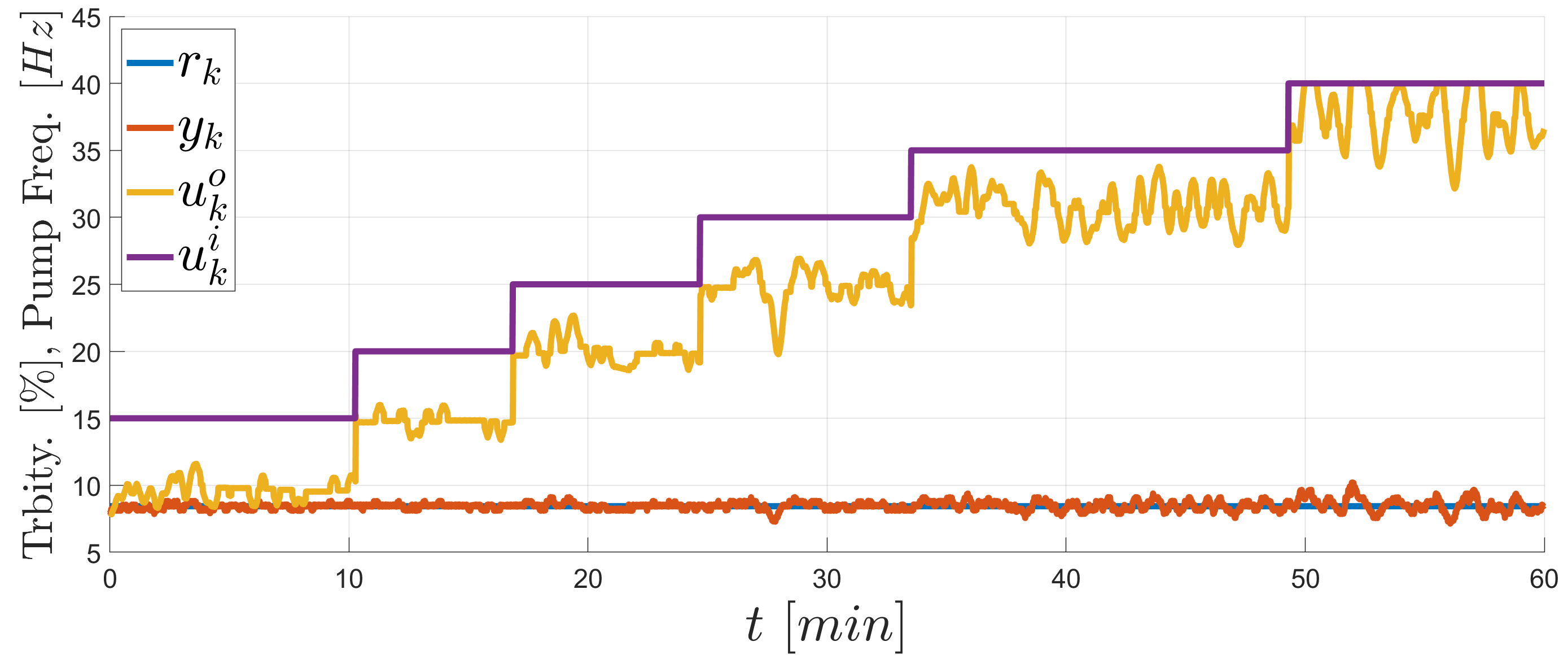}
   \caption{Closed-loop signals subject to variations of the disturbance $u^i_k$.}
   \label{Fig:ClosedLoop5}
\end{figure}

Finally, Figure \ref{Fig:ClosedLoop5}  shows the output rejection to a perturbation. In this case, the perturbation is the input flow. We can see that the turbidity remains close to the desired value ($y_k  \approx 9$) and as soon as the input flow increases, the variations of the input and output are bigger.  Due to the high input flow, there are more particles of the mixture going to the upper zone of the column, implying variations on the turbidity measurements and thus, requiring more efforts for the control input. However, as we can see in Figure \ref{Fig:ClosedLoop5}, for a constant perturbation, the variations are non increasing and the input and output remain bounded.  

\jesus{
\begin{remark}
Notice that the controller is tuned using an empirical affine piece-wise model with delay. This means that stability and performances are locally with respect to some nonlinear model describing the system. This implies that as soon as the variables move away from the linearisation point, the performances differ from the desired ones and can turn the closed-loop unstable. For instance, in Fig. \ref{Fig:ClosedLoop5}, we can see that as soon as the perturbation signal increases, the closed-loop system becomes less damped.
\end{remark}
}


%
%
%

\section{Conclusion}
{ We have proposed a simple empirical model to describe the turbidity dynamics at the top of a sedimentation column as a piece-wise time-delay linear model considering the input and output flows. The experimental results show that this model can adequately represents the dynamic of the process. In this model, all the plant parameters, including the time delay, switch between two modes. Due to the delay dependency on the switching function, we have provided sufficient conditions to guarantee closed-loop stability based on a Lyapunov-Krasovskii functional. A systematic tuning methodology for linear PI controller was proposed and validated in a pilot plant. The experimental results show the effectiveness of the proposed approach to regulate the turbidity under a wide range of operational conditions. \jesus{The proposed approach can be extended to formally deal with uncertainties, control saturation and anti-windup issues \citep{Tarbouriech2011BookStability}.}}

\small
\bibliographystyle{IEEEtran}
\bibliography{References}           

\appendix
\subsection{Proof of Proposition \ref{Prop:Main1}}\label{Proof1}
{The proof is generalized for the cases in which $d_2 \geq 1$. We consider a null reference ({\it i.e.,} $r_k = 0$ for all $k$) and \discussion{since the values $c_1$ and $c_2$ do not affect closed-loop stability, for simplicity we omit them.}} For the closed-loop system \eqref{Eq:Model}-\eqref{Eq:ControlLaw}, we define the augmented state and output vectors $x^a_k \in \mathbb{R}^{2}$ and $y^a_k \in \mathbb{R}^{2}$ respectively as:
\begin{equation} \label{Eq:AugmentedVectors}
x^a_k=\begin{pmatrix}
 x_k  \\ 
 \sum_{i=0}^{k-1}x_{i}
 \end{pmatrix}, \quad y^a_k=\begin{pmatrix}
 y_k  \\ 
 \sum_{i=0}^{k-1}y_i
 \end{pmatrix}.
\end{equation}
{Note that, since $y_k = x_k$, the augmented vectors preserve the equality $y_k^a = x_k^a $.} Then, the augmented system is written as
\begin{equation} \label{Eq:AugmentedSys}
x ^a_{k+1}=  A_{ \sigma }{x}^a_k+{ B }_{ \sigma }{u}_{k - d_{ \sigma}},\quad 
y^a_k={C}\,{x}^a_k, 
\end{equation}
and the PI controller as:
\begin{equation} \label{Eq:OutputControlLaw}
u_k =-K_c y^a_k , \quad K_c = \begin{pmatrix} k_p & k_i \end{pmatrix}. \\
\end{equation}
Since \eqref{Eq:AugmentedSys} has different delays $d_1$ and $d_2$, we select the smallest one ($d_2$ is considered the smallest one). Then, we define $h := d_1-d_2$
and the closed-loop state vector as:
\begin{equation} \label{Eq:zeta}
z_k= \begin{pmatrix}
(x^a_k) ^\top &
(x^a_{k-1})^\top &
\cdots &
(x^a_{k-d_2})^\top
\end{pmatrix}^\top.
\end{equation}
{Note that, in the considered process of this paper (with $d_2 = 1$, the closed-loop state vector is of size $4$).}
Then, the closed-loop system \eqref{Eq:AugmentedSys}-\eqref{Eq:OutputControlLaw} can be written as $z_{k+1}  = f_\sigma(z_k,z_{k-h})$ with:
\begin{equation} \label{Eq:fv} 
\begin{split}
f_1(z_k,z_{k-h}) &= \widetilde { A } _1 z_k+ A_p(z_k-z_{k-h})\\
f_2(z_k) &= \widetilde { A } _2 z_k
\end{split}
\end{equation}
\[
\widetilde{A}_{ i }=\left(\begin{smallmatrix}
{A}_{ i } & 0_2 & \cdots & 0_2 & -{B}_{ i }K_c{C}  \\ 
 I_2 & 0_2 & \cdots &0_2 &0_2 \\
0_2 & I_2 & \cdots &0_2 &0_2 \\
 \vdots & \vdots & \ddots &\vdots &\vdots \\
 0_2 & 0_2 & \cdots &I_2 &0_2 \\
 \end{smallmatrix}\right), 
{A}_{ p } = \left(\begin{smallmatrix}
0_2 & 0_2 & \cdots & 0_2 & {B}_{ 1 }K_c{C}  \\ 
 0_2 & 0_2 & \cdots &0_2 &0_2 \\
0_2 & 0_2 & \cdots &0_2 &0_2 \\
 \vdots & \vdots & \ddots &\vdots &\vdots \\
 0_2 & 0_2 & \cdots &0_2 &0_2 \\
 \end{smallmatrix} \right).
\]
with $i = \lbrace 1,2\rbrace$. Note that, for the case of $d_2 = 1$, the matrices $\widetilde{A}_1$, $\widetilde{A}_2$, and $\widetilde{A}_p = - A_p $ are equivalently represented in Proposition \ref{Prop:Main1}. Similar to \citep{Chen2003JournalDelay}, we use a common candidate Lyapunov-Krasovskii function defined as:
\begin{equation} \label{Eq:V(z)}
\begin{split}
&V_k  =V^1_k+V^2_k+V^3_k, \quad 
V^1_k ={ z }^{ T }_k P_1 z_k,\\
&V^2_k=\sum _{ \theta =-h+1 }^{ 0 }{ \sum _{ l=k-1+\theta  }^{ k-1 }{ { w }^{ T }_l{ S }_{ 1 } w_l }  } ,\;
V^3_k=\sum _{ l=k-h }^{ k-1 }{ { z }^{ T }_l{ S }_{ 2 } z_l },
\end{split}
\end{equation}
with $w_k=z_{k+1}-z_k$ and matrices
\begin{equation} \label{eq:LMI1}
P_1=P_1^T>0, \quad S_1=S_1^T>0, \quad S_2=S_2^T>0.
\end{equation}
%
We show that the CLKF is decreasing in both conditions ($\sigma = 1$ and $\sigma =2$). The difference equation $\Delta V_k$ writes:
\begin{equation}\label{Eq:DeltaV}
\Delta V_k =    V_{k+1} - V_k =  \Delta V^1 _k  + \Delta V^2 _k  + \Delta V^ 3 _k.  
\end{equation}
In the following, we find necessary conditions to guarantee $\Delta V_k<0$ for both conditions.

\subsubsection{Conditions for $\sigma=1$}

Analyzing each term
\begin{align}
&\Delta V^1_k =  2{ z }^{ T }_kP_1{ w }_k+{ w }^{ T }_kP_1{ w }_k,\\
&\Delta V^2_k =  { w }^{ T }_kh{ S }_{ 1 }{ w }_k-\sum _{ l=k-h }^{ k-1 }{ { w }^{ T }_l{ S }_1{ w } _l  }, \\
&\Delta V^3_k = { z }^{ T }_k{ S }_{ 2 }{ z }_k -{ z }^{ T }_{k-h}{ S }_{ 2 }{ z }_{k-h}. 
\end{align}
The first term of $\Delta V^1_k$ can be rewritten as 
\begin{equation}\label{eq:2XPy}
2{ z }^{ T }_k{ P }_{ 1 }w_k=
2{ \eta  }^{ T }_k{ P }^{ T }\left[ \begin{matrix} w_k \\ 0  \end{matrix} \right]
\end{equation}
where
$
\eta _k=\left[ \begin{matrix} z_k \\ w_k \end{matrix} \right] $ and $
P=\left[ \begin{matrix} { P }_{ 1 } & 0  \\ { P }_{ 2 } & { P }_{ 3 } \end{matrix} \right] 
$, with ${ P }_{ 2 }$ and ${ P }_{ 3 }$ arbitrary matrices of appropriate dimensions. From the equation (\ref{Eq:fv}) with $\sigma=1$, it follows that $0 = (\widetilde{A}_1-I){z}_k-w_k+{ A }_{ p }\displaystyle\sum _{ l=k-h }^{ k-1 }{{w}_l  }. $
Therefore, we can write (\ref{eq:2XPy}) as
\begin{equation}\label{eq:2nP}
\begin{split}
2{ \eta  }^{ T }_k{ \Pm }^{ T }\left[ \begin{matrix} \wv_k \\ \Zm  \end{matrix} \right]
& =  2{ \eta  }^{ T }_k{ \Pm }^{ T } \left[ \begin{matrix} \wv_k \\ (\widetilde{ \Am }_1-\im)\zv_k-\wv_k \end{matrix} \right] \\
& -2{ \eta  }^{ T }_k{ \Pm }^{ T }\displaystyle\sum _{ l=k-h }^{ k-1 }{ \left[ \begin{matrix} \Zm \\ \widetilde{ \Am }_{p } \end{matrix} \right] \wv_l }   \\
\end{split}
\end{equation}
\normalsize
where $\widetilde{\Am}_p=-{\Am_p}$. The second term of the equation (\ref{eq:2nP}) can be bounded above by using the Moon's Inequality as in  \citep{Chen2003}
\begin{align*}
&-2\displaystyle\sum _{ l=k-h }^{ k-1 }{ { \eta  }^{ T }_k{ \Pm }^{ T }\left[ \begin{matrix} \Zm  \\ \widetilde{ \Am }_{ p } \end{matrix} \right] \wv_l } \\  &\le \displaystyle\sum _{ l=k-h }^{ k-1 }{ { \left[ { \begin{matrix} \eta _k \\ \wv_l \end{matrix} } \right]  }^{ T }\left[ \begin{smallmatrix} \Wm & \Mm-{ \Pm }^{ T }\left[ \begin{matrix} \Zm  \\ \widetilde{ \Am }_{ p } \end{matrix} \right]  \\ { \Mm }^{ T }-\left[ \begin{matrix} \Zm  & \widetilde{ \Am }_{ p }^{ T } \end{matrix} \right] \Pm & { \Sm }_{ 1 } \end{smallmatrix} \right]  } \left[ { \begin{matrix} \eta _k \\ \wv_l \end{matrix} } \right],\\ 
&=h{ \eta  }^{ T }_k\Wm\eta _k + 2{ \eta  }^{ T }_k\left( \Mm-{ \Pm }^{ T }\left[ \begin{matrix} \Zm  \\ { \Am }_{ d } \end{matrix} \right]  \right) \left( \zv_k-\zv_{k-h} \right) \\
& +\displaystyle\sum _{ l=k-h }^{ k-1 }{{\wv}^T_l\Sm_1{\wv}_l},
\end{align*}
\begin{equation}\label{eq:LMI4}
\left[ \begin{matrix} \Wm & \Mm \\ { \Mm }^{ T } & { \Sm }_{ 1 } \end{matrix} \right] \ge 0, \;
\Wm= \begin{bmatrix} { \Wm }_{ 1 } & { \Wm }_{ 2 } \\ { \Wm }_{ 2 }^{ T } & { \Wm }_{ 3 } \end{bmatrix}, \;
\Mm= \left[ \begin{matrix} { \Mm }_{ 1 } \\ { \Mm }_{ 2 } \end{matrix} \right] .
\end{equation}
Finally, one can bound (\ref{Eq:DeltaV}) by
\begin{align*}
\Delta V_k\le
& \,\,\, h{ \eta  }^{ T }_k\Wm\eta _k +{ \zv }^{ T }_k{ \Sm }_{ 2 }\zv_k\\
&\hspace{-1cm}+2{ \eta  }^{ T }_k\left( \Mm-{ \Pm }^{ T }\left[ \begin{smallmatrix} \Zm  \\ \widetilde{ \Am }_{ p } \end{smallmatrix} \right]  \right) \left( \zv_k-\zv_{k-h} \right)  +{ \wv }^{ T }_k\left[ { \Pm }_{ 1 }+h{ \Sm }_{ 1 } \right] \wv_k\\
&\hspace{-1cm} +2{ \eta  }^{ T }_k{ \Pm }^{ T }\left[ \begin{matrix} \wv_k \\ \left( \widetilde{\Am}_1-\im \right) \zv_k-\wv_k \end{matrix} \right]  -{ \zv }^{ T }_{k-h}{ \Sm }_{ 2 }\zv_{k-h},
\end{align*}
which can be expressed as $\Delta V_k\le { \overline{\zv}_k  ^ T }\Lambdam \overline{\zv} _k,
$
with the new state vector $\overline{\zv}_k = \left[ \begin{smallmatrix} \eta _k \\ \zv_{k-h} \end{smallmatrix} \right] =\left[ \begin{smallmatrix} \zv_k \\ w_k \\ \zv_{k-h} \end{smallmatrix} \right] $
and the matrix $\Lambdam$ defined in the statement of Proposition \ref{Prop:Main1}. Therefore, if 
\begin{equation}\label{eq:LMI5}
 \Lambdam  < 0,
\end{equation}
then $\Delta V_k < 0 $.

\subsubsection{Conditions for $\sigma=2$}

Analysing each term
\begin{align*}
&\Delta V^1_{k} =  { \zv }^{ T }_{k}\left[\widetilde\Am^T_2\Pm_1\widetilde\Am_2-\Pm_1  \right]\zv _{k},\\
&\Delta V^2_{k} =  { \wv }^{ T }_{k}h{ \Sm }_{ 1 }{ \wv }_{k}-\sum _{ l=k-h }^{ k-1 }{ { \wv }^{ T }_{l}{ \Sm }_1{ \wv } _{l}  },\\ 
&\Delta V^3_{k} = { \zv }^{ T }_{k}{ \Sm }_{ 2 }{ \zv }_{k} -{ \zv }^{ T }_{k-h}{ \Sm }_{ 2 }{ \zv }_{k-h}. 
\end{align*}
{Note that, for $\sigma = 2$ the dynamic is given by $z_{k+1} = \widetilde{A}_2 z_k$ and thus $w_k = (\widetilde{A}_2 - I)z_k$. Then, the second term of the Lyapunov difference equation can be written as:
\begin{align*}
\Delta V_k ^2  &= z_{k-h}^T \left(h\widetilde{A}_2^{h^T} \tilde{S}_1 \widetilde{A}_2^{h} - \sum _{i = 0} ^{h-1} \widetilde{A}_2 ^{i ^ T} \tilde{S}_1 \widetilde{A}_2 ^{i}\right)z_{k-h}.
\end{align*}
with $\tilde{S}_1 = (\widetilde{A}_2 - I)^T S_1 (\widetilde{A}_2 - I)$.
Similarly the $\Delta V^3_{k}$ term is written in terms of the delayed variable $\zv _{k-h}$ as follows:
\begin{align*}
\Delta V^3_{k} =& { \zv }^{ T }_{k}{ \Sm }_{ 2 }{ \zv }_{k} -{ \zv }^{ T }_{k-h}{ \Sm }_{ 2 }{ \zv }_{k-h},\\
=& { \zv }^{ T }_{k-h}  (\widetilde\Am^T_2)^h { \Sm }_{ 2 }(\widetilde\Am_2)^h{ \zv }_{k-h} 
 -{ \zv }^{ T }_{k-h}{ \Sm }_{ 2 }{ \zv }_{k-h}.
\end{align*}
Combining $\Delta V_k ^2 $ and $\Delta V_k ^3$, we obtain:
\begin{equation*}
\begin{split}
 \Delta V_k ^3  + \Delta V_k ^2 & = z_{k-h}^T \left((\widetilde\Am^T_2)^h { \Sm }_{ 2 }(\widetilde\Am_2)^h 
 -{ \Sm }_{ 2 }  \right)z_{k-h}\\
&\hspace{-0.5cm} + z_{k-h}^T \left( h (\widetilde{A}_2^{T})^h \tilde{S}_1 (\widetilde{A}_2)^{h} - \sum _{i = 0} ^{h-1} \widetilde{A}_2 ^{i ^ T} \tilde{S}_1 \widetilde{A}_2 ^{i}  \right)z_{k-h} .
 \end{split}
\end{equation*}
Finally, if $\Delta V_k^1<0$ and $\Delta V _k^2 + \Delta V_k^3<0$, one can guarantee $\Delta V_k <0$. This leads respectively to the following LMI conditions:
\begin{equation}\label{eq:LMI6}
\widetilde\Am^T_2\Pm_1\widetilde\Am_2-\Pm_1    
 < 0, \quad \Lambda_2 < 0
\end{equation}
with $\Lambda_2 = h (\widetilde{A}_2^{T})^h \tilde{S}_1 (\widetilde{A}_2)^{h} - \sum _{i = 0} ^{h-1} \widetilde{A}_2 ^{i ^ T} \tilde{S}_1 \widetilde{A}_2 ^{i}  +(\widetilde\Am^T_2)^h { \Sm }_{ 2 }(\widetilde\Am_2)^h 
 -{ \Sm }_{ 2 } $. 
}

Summarizing, the stability of the closed-loop system is stablished if there exist matrices $\Pm_1$, $\Pm_2$, $\Pm_3$, $\Sm_1$, $\Sm_2$, $\Wm_1$, $\Wm_2$, $\Wm_3$, $\Mm_1$ and $\Mm_2$ such that LMIs (\ref{eq:LMI1}), (\ref{eq:LMI4}), (\ref{eq:LMI5}), (\ref{eq:LMI6}) are satisfied.


%
%


 \subsection{Proof of Proposition \ref{Prop:Main2}}\label{Proof2}
If the LMI \eqref{Ineq:LMIS2} is satisfied, then for both $\sigma=1$ and $\sigma=2$ the variation of the CLKF is given by $\Delta V_{k} = V_{k+1}-V_{k} < -{z}^T_{k}{\Qm}{z}_{k} $. By adding at both sides from $k=0$ to $k=\infty$, one can rebuild the cost function $J$ in \eqref{Eq:CostFunction} as $\sum _{ i=0 }^{ \infty  } \left(V_{i+1}-V_{i}\right)   < -\sum _{ i=0 }^{ \infty  }{\zv}^T_{i}\Qm{\zv}_{i} = -J$. Then, by computing the right hand side and re-organizing terms we obtain $J < V_0 - V_\infty.$
Since the closed-loop system is asymptotically stable (Proposition \ref{Prop:Main1}), $V_\infty=0$. Then, considering that $z_k=z_0$, $\forall k \in [-h, 0]$, and evaluating $V_0$ in \eqref{Eq:V(z)}, we obtain the following bound for the cost function 
\begin{equation}
\label{eq:P1+hS2}
\begin{split}
J < V_0=\zv^T_0(\Pm_1+h\Sm_2)\zv_0.\\
\end{split}
\end{equation}
Finally, the minimization of $J$ in \eqref{Eq:CostFunction} is equivalent to the minimization of the maximum eigenvalue of $\Pm_1+h\Sm_2$.


\end{document}